\documentclass[12pt,a4paper]{article}

\usepackage{epsfig,graphicx}
\usepackage{amsmath,amsfonts,amssymb}
\usepackage{citesort}
\newcommand{\unit}{\leavevmode\hbox{\small1\kern-3.6pt\normalsize1}}

\def \GeV{{\mathrm{GeV}}}
\def \TeV{{\mathrm{TeV}}}

\newcommand{\mdm}{{m_{\textrm{DM}}}}
\newcommand{\str}{\,\textrm{str}}

\newcommand{\Rsun}{{R_{\odot}}}
\def\tanb{\tan\beta}

\def\lsim{\raise0.3ex\hbox{$\;<$\kern-0.75em\raise-1.1ex\hbox{$\sim\;$}}}
\def\gsim{\raise0.3ex\hbox{$\;>$\kern-0.75em\raise-1.1ex\hbox{$\sim\;$}}}
\newcommand{\dis}[1]{\begin{equation}\begin{split}#1\end{split}\end{equation}}

\newcommand{\be}{\begin{equation}}
\newcommand{\mpc}{\,\textrm{Mpc}}

\def    \bea           	{\begin{eqnarray}}
\def    \eea           	{\end{eqnarray}}

\newcommand\treh{T_{\rm R}}
\newcommand{\tev}{\,\textrm{TeV}}
\newcommand{\gev}{\,\textrm{GeV}}

\newcommand{\km}{\,\textrm{km}}
\newcommand{\cm}{\,\textrm{cm}}

\newcommand{\photino}{\tilde{\gamma}}

\newcommand{\mgluino}{{m_{\tilde{g}}}}
\newcommand{\taudm}{{\tau_{\textrm{DM}}}}

\newcommand{\bfrac}[2]{{\left(\frac{#1}{#2} \right)  }}

\begin{document}

\thispagestyle{empty}
\begin{flushright}
%DESY 04-129\\
%IFIC/04-44\\
FTUAM 09/15\\
IFT-UAM/CSIC-09-29\\
%hep-ph/yymmddd\\
\vspace*{5mm}{June 2009}
\end{flushright}

\begin{center}
{\Large \textbf{ 
\bf 
%Gravitino dark matter in the  $\mu\nu$SSM
Gamma-ray detection from gravitino dark matter decay in the $\mu\nu$SSM
} 
}

\vspace{0.5cm}
\hspace*{-1mm}
Ki-Young Choi$^{a,b}$, Daniel E.~L\'opez-Fogliani$^{c}$, Carlos Mu\~noz$^{a,b}$ and Roberto Ruiz de Austri$^d$\\[0.4cm] 
{$^{a}$\textit{Departamento de F\'{\i}sica Te\'{o}rica, Universidad Aut\'{o}noma de Madrid,\\
Cantoblanco, E-28049 Madrid, Spain}}\\[0pt]
{$^{b}$\textit{Instituto de F\'{\i}sica Te\'{o}rica UAM/CSIC, Universidad Aut\'{o}noma de Madrid,\\
Cantoblanco, E-28049 Madrid, Spain}}\\[0pt]
{$^{c}$\textit{Department of Physics and Astronomy, University of Sheffield,\\ Sheffield S3 7HR, England}}\\[0pt]
{$^{d}$\textit{Instituto de F\'{\i}sica Corpuscular UV/CSIC, Universidad de Valencia,\\ Edificio Institutos de Paterna, Apt. 22085, E-46071 Valencia, Spain}}\\[0pt]

\begin{abstract}
The $\mu\nu$SSM provides a solution to the
$\mu$-problem of the MSSM and explains the origin of neutrino masses
by simply using right-handed neutrino superfields. 
Given that R-parity is broken in this model, the gravitino is a natural candidate for dark matter since its lifetime becomes much longer than the age of the Universe.
We consider the implications of gravitino dark matter in the $\mu\nu$SSM,
analyzing in particular the prospects for detecting gamma rays 
from decaying gravitinos. 
If the gravitino explains the whole dark matter component,
a gravitino mass larger than $20\gev$ is disfavored by 
the isotropic diffuse photon background measurements. 
On the other hand, a gravitino with a mass range between $0.1 - 20$ GeV gives rise to 
a signal that might be observed by the FERMI satellite. 
In this way important regions of the parameter space of the $\mu\nu$SSM can be
checked.
%important regions of the parameter space of the $\mu\nu$SSM can be checked in this way.

% ---------
% We consider the implications of gravitino dark matter in this model
% from the collider and cosmological effects with the possbile range of 
%  gravitino mass.
\end{abstract}
\end{center}

%\vspace*{15mm}\hspace*{3mm}
%{\small PACS}: 12.60.Jv, 14.60.St 
%95.35.+d  ?????????? % Supersymmetric models, Dark matter
\newpage

\section{Introduction}\label{Introduction}

The
``$\mu$ from $\nu$'' Supersymmetric Standard Model
($\mu\nu$SSM) 
was proposed in the literature \cite{MuNuSSM,MuNuSSM2,MuNuSSM0}
as an alternative to the Minimal Supersymmetric Standard Model (MSSM). In particular, it provides a solution to the
$\mu$-problem \cite{muproblem} of the MSSM and explains the origin of neutrino masses
by simply using right-handed neutrino superfields.
%relying on the

The superpotential of the $\mu\nu$SSM contains, in addition to the
usual Yukawas for quarks and charged leptons,
Yukawas for neutrinos
$\hat H_u\,  \hat L \, \hat \nu^c$, terms of the type
$\hat \nu^c \hat H_d\hat H_u$ 
producing an effective  $\mu$ term through right-handed sneutrino
vacuum expectation values (VEVs),
and also terms of the type $\hat \nu^c \hat \nu^c \hat \nu^c$  
avoiding the existence of a Goldstone boson and
contributing to generate
effective Majorana masses for neutrinos at the electroweak scale.
Actually, the explicit breaking of R-parity in this model 
by the above terms produces the mixing of neutralinos with
left- and right-handed neutrinos, and as a consequence a generalized matrix of the
seesaw type that gives rise at tree level to three
light eigenvalues corresponding to neutrino 
masses \cite{MuNuSSM}.

% The possibility as a source of baryon asymmetry was studied
% in~\cite{Farzan:2005ez}. 

The breaking of R-parity can easily be understood if we realize that in the limit 
where Yukawas for neutrinos are vanishing, the 
$\hat \nu^c$ are just ordinary singlet superfields, 
without any connection with neutrinos,
and
this model would coincide (although with three instead of one singlet) with the
Next-to-Minimal Supersymmetric Standard Model (NMSSM) where R-parity is 
conserved.
Once we switch on the neutrino Yukawa couplings, the fields
$\hat \nu^c$ become right-handed neutrino superfields, and, as a consequence, R-parity
is broken. Indeed this breaking is small because, as mentioned above, we have an electroweak-scale seesaw, implying neutrino Yukawa couplings no larger than $10^{-6}$
(like the electron Yukawa).

The latter also implies that processes violating lepton number that might wash-out any baryon asymmetry present in the model would be suppressed.
Notice also that electroweak baryogenesis could work in this model in a similar way to the case of the
NMSSM \cite{teixeira}. 
Actually, the fact that in the $\mu\nu$SSM there are three singlets instead of one like in the NMSSM, should in principle give more freedom to be able to obtain more easily electroweak baryogenesis. 
The detail conditions for baryogenesis in this model are presently under study \cite{chung}.

Since R-parity is broken in the $\mu\nu$SSM, one could worry about fast proton decay through the usual baryon and lepton number violating operators of the MSSM.
Nevertheless, the choice of $R$-parity is {\it ad hoc}. There are other discrete 
symmetries, 
like e.g. baryon triality which only forbids the baryon violating operators \cite{dreiner3}.
Obviously, for all these symmetries R-parity is violated.
Besides, in string constructions the matter superfields can be located in different sectors of the compact space or have different extra $U(1)$ charges, in such a way that 
some operators violating $R$-parity can be forbidden \cite{old}, 
but others can be allowed.

Several recent papers have studied different aspects of the $\mu\nu$SSM.
In \cite{MuNuSSM2},
the parameter space of the model was analyzed
in detail, studying the viable regions which avoid false minima and tachyons, as well as fulfill the Landau pole constraint. The structure of the mass matrices, and the
associated particle spectrum was also computed, paying
special attention to the mass of the lightest Higgs.
In \cite{Ghosh:2008yh}, neutrino masses and mixing angles were discussed, as well as the decays of the lightest neutralino
to two body ($W$-lepton) final states.
The correlations of the decay branching ratios with the neutrino mixing angles were studied as another possible test of the $\mu\nu$SSM at the LHC.
%Similar issues were studied
The phenomenology of the $\mu\nu$SSM was also studied
in \cite{Hirsch0}, particularized for one and two generations
of right-handed sneutrinos, and taking into account all 
possible final states when studying the decays of the lightest neutralino. 
Possible signatures that might allow to distinguish this model from other R-parity breaking models were discussed qualitatively in these two works \cite{Ghosh:2008yh,Hirsch0}. 
% Let us finally mention that
% terms of the type
% $\hat \nu^c \hat H_d \hat H_u$ and $\hat \nu^c \hat \nu^c \hat \nu^c$
% were also analysed as sources of the observed baryon asymmetry
% in the Universe \cite{vallle} and of neutrino masses and bilarge mixing \cite{sri}, respectively.
In \cite{neutrinos},
the analysis of the vacua of the $\mu\nu$SSM carried out in \cite{MuNuSSM2} was completed, obtaining that spontaneous CP violation through complex Higgs and sneutrino
VEVs is possible. Neutrino physics and the associated electroweak seesaw mechanism was also studied. It was shown how the experimental results can easily 
be reproduced and explained why the mixing patterns are so different in the
quark and lepton sectors. All the results were discussed
in the general case with phases.

On the other hand, 
when $R$-parity is broken, the lightest supersymmetric particle (LSP) is no longer stable. Thus
neutralinos \cite{reviewmio} or sneutrinos \cite{sneutrino}, with
very short lifetimes,
are no longer candidates for the dark matter of the Universe.
Nevertheless, if the gravitino is the LSP its decay is suppressed both by
the gravitational interaction and by the small R-parity violating coupling, and as a 
consequence its lifetime can be much longer than the age of the 
Universe \cite{Takayama:2000uz}.
Thus the gravitino can be in principle a dark matter candidate in R-parity breaking models.
This possibility and its phenomenological consequences were studied mainly in the context of bilinear or trilinear R-parity violation 
scenarios in \cite{Takayama:2000uz,Hirsch,buchmuller,Lola,bertone,ibarra,tran,moro,covi}.
In \cite{buchmuller,bertone,ibarra,moro} the prospects for detecting gamma rays
from decaying gravitinos
in satellite experiments were also analyzed.
In this work we want to discuss these issues, gravitino dark matter and
its possible detection in the FERMI satellite \cite{Fermi0}, in the context of the $\mu\nu$SSM.

% [We will consider the collider and cosmological implications of this model
% with Gravitino as dark matter]

\section{The $\mu\nu$SSM}\label{parameters}

The  superpotential of the $\mu \nu$SSM introduced in \cite{MuNuSSM} is 
given by
\begin{align}\label{superpotential}
W = &
\ \epsilon_{ab} \left(
Y_{u_{ij}} \, \hat H_u^b\, \hat Q^a_i \, \hat u_j^c +
Y_{d_{ij}} \, \hat H_d^a\, \hat Q^b_i \, \hat d_j^c +
Y_{e_{ij}} \, \hat H_d^a\, \hat L^b_i \, \hat e_j^c +
Y_{\nu_{ij}} \, \hat H_u^b\, \hat L^a_i \, \hat \nu^c_j 
\right)
\nonumber\\
& 
-\epsilon{_{ab}} \lambda_{i} \, \hat \nu^c_i\,\hat H_d^a \hat H_u^b
+
\frac{1}{3}
\kappa{_{ijk}} 
\hat \nu^c_i\hat \nu^c_j\hat \nu^c_k \ ,
%\label{superpotential}
\end{align}
where we take $\hat H_d^T=(\hat H_d^0, \hat H_d^-)$, $\hat H_u^T=(\hat
H_u^+, \hat H_u^0)$, $\hat Q_i^T=(\hat u_i, \hat d_i)$, $\hat
L_i^T=(\hat \nu_i, \hat e_i)$, 
$i,j,k=1,2,3$ are family indices, $a,b=1,2$ are $SU(2)_L$ indices with
$\epsilon_{12}=1$, and
$Y$, $\lambda$, $\kappa$ are dimensionless
matrices, a vector, and a totally symmetric tensor, respectively.

Working in the framework of supergravity,
%gravity mediated supersymmetry breaking, 
the Lagrangian  $\mathcal{L}_{\text{soft}}$ 
is given by:
\begin{eqnarray}
-\mathcal{L}_{\text{soft}} & =&
 m_{\tilde{Q}_{ij} }^2\, \tilde{Q^a_i}^* \, \tilde{Q^a_j}
+m_{\tilde{u}_{ij}^c}^{2} 
\, \tilde{u^c_i}^* \, \tilde u^c_j
+m_{\tilde{d}_{ij}^c}^2 \, \tilde{d^c_i}^* \, \tilde d^c_j
+m_{\tilde{L}_{ij} }^2 \, \tilde{L^a_i}^* \, \tilde{L^a_j}
+m_{\tilde{e}_{ij} ^c}^2 \, \tilde{e^c_i}^* \, \tilde e^c_j
\nonumber \\
&+ &
m_{H_d}^2 \,{H^a_d}^*\,H^a_d + m_{H_u}^2 \,{H^a_u}^* H^a_u +
m_{\tilde{\nu}_{ij}^c}^2 \,\tilde{{\nu}^c_i}^* \tilde\nu^c_j 
\nonumber \\
&+&
\epsilon_{ab} \left[
(A_uY_u)_{ij} \, H_u^b\, \tilde Q^a_i \, \tilde u_j^c +
(A_dY_d)_{ij} \, H_d^a\, \tilde Q^b_i \, \tilde d_j^c +
(A_eY_e)_{ij} \, H_d^a\, \tilde L^b_i \, \tilde e_j^c 
\right.
\nonumber \\
&+&
\left.
(A_{\nu}Y_{\nu})_{ij} \, H_u^b\, \tilde L^a_i \, \tilde \nu^c_j 
+ \text{c.c.}
\right] 
\nonumber \\
&+&
\left[-\epsilon_{ab} (A_{\lambda}\lambda)_{i} \, \tilde \nu^c_i\, H_d^a  H_u^b
+
\frac{1}{3}
(A_{\kappa}\kappa)_{ijk} \, 
\tilde \nu^c_i \tilde \nu^c_j \tilde \nu^c_k\
+ \text{c.c.} \right]
\nonumber \\
&-&  \frac{1}{2}\, \left(M_3\, \tilde\lambda_3\, \tilde\lambda_3+M_2\,
  \tilde\lambda_2\, \tilde
\lambda_2
+M_1\, \tilde\lambda_1 \, \tilde\lambda_1 + \text{c.c.} \right) \,.
\label{2:Vsoft}
\end{eqnarray}
%
%one obtains from (\ref{2:Vsoft}) 
%the neutral scalars develop in general
%the VEVs:
%
%
%\begin{equation}\label{2:vevs}
%\langle H_d^0 \rangle = v_d \, , \quad
%\langle H_u^0 \rangle = v_u \, , \quad
%\langle \tilde \nu_i \rangle = \nu_i \, , \quad
%\langle \tilde \nu_i^c \rangle = \nu_i^c \,,
%\end{equation}
% 
In addition to terms from $\mathcal{L}_{\text{soft}}$, the 
tree-level scalar potential receives the $D$ and $F$ term
contributions also computed in \cite{MuNuSSM}.
In the following we will assume for simplicity that all parameters in the potential are real. 
Once the electroweak symmetry is spontaneously broken, the neutral scalars develop in general the following VEVs:
\begin{equation}\label{2:vevs}
\langle H_d^0 \rangle = v_d \, , \quad
\langle H_u^0 \rangle = v_u \, , \quad
\langle \tilde \nu_i \rangle = \nu_i \, , \quad
\langle \tilde \nu_i^c \rangle = \nu_i^c \,.
%\label{esperados}
\end{equation}

For our computation below we are interested in the neutral fermion mass matrix.
As explained in \cite{MuNuSSM,MuNuSSM2},
neutralinos mix with the neutrinos and therefore in a basis where
${\chi^{0}}^T=(\tilde{B^{0}},\tilde{W^{0}},\tilde{H_{d}},\tilde{H_{u}},\nu_{R_i},\nu_{L_i})$,
one obtains the following neutral fermion mass terms in the Lagrangian
%$\mathcal{L}_{\mathrm{neutral}}^{\mathrm{mass}}
%^{\tilde \chi^0} 
%=
\begin{equation}
-\frac{1}{2} (\chi^0)^T \mathcal{M}_{\mathrm{n}}
%_{\tilde \chi^0} 
\chi^0 + \mathrm{c.c.}\ ,
\label{matrixneutralinos}
\end{equation}
where
\begin{align}
%\textrm{\cal{M}}_n
{\cal M}_n=\left(\begin{array}{cc}
M & m\\
m^{T} & 0_{3\times3}\end{array}\right),
\label{matrizse}
\end{align}
with
{\small  \begin{align}
\hspace*{-2.5cm} \hspace{.2mm} M=\hspace{-.2mm}
\left(
\begin{array}{ccccccc}
M_{1} & 0 & -A v_{d} & A v_{u} & 0 & 0 & 0\\
0 & M_{2} &B v_{d} & -B v_{u} & 0 & 0 & 0\\
-A v_{d} & B v_{d} & 0 & -\lambda_{i}\nu^c_{i} & -\lambda_{1}v_{u} & -\lambda_{2}v_{u} & -\lambda_{3}v_{u}\\
A v_{u} & -B v_{u} & \: \: -\lambda_{i}\nu^c_{i} & 0 & -\lambda_{1}v_{d}+Y_{\nu_{i1}}\nu_{i} & -\lambda_{2}v_{d}+Y_{\nu_{i2}}\nu_{i} & -\lambda_{3}v_{d}+Y_{\nu_{i3}}\nu_{i}\\
0 & 0 &  -\lambda_{1}v_{u} & \: \:-\lambda_{1}v_{d}+Y_{\nu_{i1}}\nu_{i} & 2\kappa_{11j}\nu^c_{j} & 2\kappa_{12j}\nu^c_{j} & 2\kappa_{13j}\nu^c_{j}\\
0 & 0 & -\lambda_{2}v_{u} &  \: \: -\lambda_{2}v_{d}+Y_{\nu_{i2}}\nu_{i} & 2\kappa_{21j}\nu^c_{j} & 2\kappa_{22j}\nu^c_{j} & 2\kappa_{23j}\nu^c_{j}\\
0 & 0 & -\lambda_{3}v_{u} & \: \:-\lambda_{3}v_{d}+Y_{\nu_{i3}}\nu_{i} & 2\kappa_{31j}\nu^c_{j} & 2\kappa_{32j}\nu^c_{j} & 2\kappa_{33j}\nu^c_{j}\end{array}
\right)\ ,
\label{neumatrix}
\end{align}
}
where $A=\frac{G}{\sqrt{2}} \sin\theta_W$, $B=\frac{G}{\sqrt{2}} \cos\theta_W$,
$G^2\equiv g_{1}^{2}+g_{2}^{2}$,
and
\begin{align}
m^{T}=\left(\begin{array}{ccccccc}
-\frac{g_{1}}{\sqrt{2}}\nu_{1} \: & \: \frac{g_{2}}{\sqrt{2}}\nu_{1} & \: 0 & \: Y_{\nu_{1i}}\nu^c_{i} & \: Y_{\nu_{11}}v_{u} & \: Y_{\nu_{12}}v_{u} & \: Y_{\nu_{13}}v_{u}\\
\: -\frac{g_{1}}{\sqrt{2}}\nu_{2} & \: \frac{g_{2}}{\sqrt{2}}\nu_{2} & \: 0 & \: Y_{\nu_{2i}}\nu^c_{i} & \: Y_{\nu_{21}}v_{u} & \: Y_{\nu_{22}}v_{u} & \: Y_{\nu_{23}}v_{u}\\
\: -\frac{g_{1}}{\sqrt{2}}\nu_{3}\: & \: \frac{g_{2}}{\sqrt{2}}\nu_{3} & \: 0 & \: Y_{\nu_{3i}}\nu^c_{i} & \: Y_{\nu_{31}}v_{u} & \: Y_{\nu_{32}}v_{u} & \: Y_{\nu_{33}}v_{u}\end{array}\right)\ .\end{align}
The above $10\times 10$ matrix, Eq.~(\ref{matrizse}), is of the seesaw type giving rise 
to the neutrino masses which have to be very small. 
This is the case since the entries of the matrix $M$ are much
larger than the ones in the matrix $m$.
Notice in this respect that the entries of $M$ are of the order of the 
electroweak scale while the ones in $m$ are of the order 
of the Dirac masses for the neutrinos \cite{MuNuSSM,MuNuSSM2}.

At low energy the free parameters of the $\mu\nu$SSM in the neutral scalar sector
are \cite{MuNuSSM2}: 
$\lambda_i$, $\kappa_{ijk}$, $m_{H_d}$, $m_{H_u}$, $m_{\widetilde{L}_{ij}}$, 
$m_{\widetilde{\nu}_{ij}^{c}}$,
$A_{\lambda_i}$, $A_{\kappa_{ijk}}$, and $A_{\nu_{ij}}$.
Strong upper bounds upon the intergenerational scalar mixing 
exist, so in the following we assume that 
such mixings are negligible, and therefore the sfermion soft mass 
matrices are diagonal in the flavour space.
Thus using the eight minimization conditions for the neutral scalar potential,
one can eliminate the soft masses $m_{H_d}$, $m_{H_u}$, $m_{\widetilde{L}_{i}}$, and $m_{\widetilde{\nu}_{i}^{c}}$
in favour of the VEVs
$v_d$, $v_u$,
$\nu_i$, and $\nu^c_i$.
On the other hand, using the Standard Model Higgs VEV,
$v\approx 174$ GeV, 
$\tan\beta$, and $\nu_i$, one can determine the SUSY Higgs 
VEVs, $v_d$ and $v_u$, through $v^2 = v_d^2 + v_u^2 + \nu_i^2$. 
We thus consider as independent parameters
the following set of variables:
\bea
\lambda_i, \, \kappa_{ijk},\, \tan\beta, \, \nu_i, \nu^c_i, \, A_{\lambda_i}, \, A_{\kappa_{ijk}}, \, A_{\nu_{ij}}\ .
\label{freeparameters}
\eea
It is worth remarking here that, because of the minimization conditions,
the VEVs of the left-handed sneutrinos,
%  $\tilde \nu$, 
$\nu_i$, are in general small, of the order of Dirac masses for the 
neutrinos \cite{MuNuSSM}.
Then, since $\nu_i<<v_d, v_u$ we can define the above value of $\tanb$ as usual,
$\tan\beta=\frac{v_u}{v_d}$.

We will assume for simplicity that there is no intergenerational mixing
in the parameters of the model,
%eq. (\ref{freeparameters}), 
and that in general they have
the same values for the three families.
In the case of neutrino parameters, 
following the discussion in \cite{neutrinos,MuNuSSM2}, 
we need at least two generations with different VEVs and couplings in order to obtain the
correct experimental pattern. We choose 
$Y_{\nu_1} \neq Y_{\nu_2}= Y_{\nu_3}$ and $\nu_1\neq \nu_2=\nu_3$.
Thus the low-energy free parameters in our analysis are
\bea
\lambda, \, \kappa,\, \tan\beta, \, \nu_1, \,  \nu_3, \nu^c, \, A_{\lambda}, \, A_{\kappa}, \, A_\nu\ ,
\label{freeparameters2}
\eea
%where we have chosen $\nu_1\neq  \nu_2=\nu_3$, and 
where we have defined ${\lambda} \equiv {\lambda_i}$, 
$\kappa\equiv {\kappa_{iii}}$, 
$\nu^c \equiv \nu^c_i$, $A_{\lambda} \equiv A_{\lambda_i}$, 
$A_{\kappa} \equiv A_{\kappa_{iii}}$,
$A_{\nu} \equiv A_{\nu_{ii}}$.
Actually, we have checked that with
$Y_{\nu_2}= Y_{\nu_3} \approx 2 \; Y_{\nu_1} \sim 10^{-6} $ and
$\nu_2=\nu_3 \approx 2 \, \nu_1 \sim 10^{-4}$~GeV, the observed neutrino masses and mixing angles are reproduced. As explained in detail in \cite{neutrinos}, 
this result is obtained so easily due to the peculiar characteristics of this seesaw, where R-parity is broken and the relevant scale is the electroweak scale.

% Following \cite{neutrinos} we choose diagonal Yukawa couplings for neutrinos with$Y_{\nu_1} \neq Y_{\nu_2}= Y_{\nu_3}$ , that also with $\nu_1\neq \nu_2=\nu_3$ gives degenerative  $\nu_\mu-\nu_{\tau}$, that easy reproduce the experimental patterns.

The soft SUSY-breaking terms, namely gaugino masses,
$M_{1,2,3}$, scalar masses, $m_{\tilde Q,\tilde u^c,\tilde d^c,\tilde e^c}$,
and trilinear parameters, $A_{u,d,e}$, are also
taken as free parameters and specified at low scale.
% Data on neutrino masses, and the usual 
% Standard Model parameters such as fermion and gauge boson masses,
% the fine structure constant $\alpha (M_Z)$, the Fermi constant from 
% muon decay $G_F^\mu$, and the strong coupling constant $\alpha_s(M_Z)$,  
% will be used in the computation~\cite{pdg07}. Concerning the top mass,
% we will take $m_t = 172.6$ GeV~\cite{topmass:mar08}.

\section{Gravitino dark matter}

%\subsection{Relic density of Gravitino}

Let us now show that the lifetime of the gravitino LSP is typically much longer than
the age of the Universe in the $\mu\nu$SSM, and therefore it can be in principle a candidate for dark matter.
In the supergravity Lagrangian there is an interaction term between
the gravitino, the field strength for the photon, and the photino. 
Since, as discussed above, due to the breaking of R-parity the photino and the left-handed neutrinos are mixed, the gravitino will be able to decay through the
interaction term into a photon
and a neutrino \cite{Takayama:2000uz}\footnote{Other possible decay modes such as gravitino decay into a $W^{\pm}$ and a charged lepton,
or into a $Z^0$ and a neutrino \cite{ibarra} are not relevant in our case, since we will obtain below that a gravitino  mass
smaller than 20 GeV is convenient in order to fulfill
experimental constraints. Neither we consider the possibility that
the gravitino might in principle decay to singlet Higgs-neutrino if the
Higgs is sufficiently light.}. 
Thus one obtains:
\bea
\Gamma(\Psi_{3/2}
%\widetilde{G} 
\to \sum_i \gamma \nu_i)\simeq  \frac{1}{32 \pi} \mid U_{\widetilde{\gamma} \nu} \mid^2 \frac{m^3_{3/2}}{M_P^2} \label{c1}
\ ,
\eea
where $m_{3/2}$ is the gravitino mass, $M_P=2.4\times 10^{18} \gev$ is the reduced Planck mass, and 
$|U_{\widetilde{\gamma}\nu}|^{2}$ determines the photino content of the neutrino
\begin{equation}
|U_{\widetilde{\gamma}\nu}|^{2}=\sum_{i=1}^{3}|N_{i1}\cos\theta_{W}+
N_{i2}\sin\theta_{W}|^{2}\ .
\label{umix}
\end{equation}
Here $N_{i1}$ ($N_{i2}$) is the Bino (Wino) component of the $i$-neutrino.

The lifetime of the gravitino can then be written as
\dis{
\tau_{3/2} \simeq 3.8\times 10^{27}\ \text{s}
\left( \frac{\left| U_{\photino \nu}  \right|^2}{10^{-16}} \right)^{-1}
\left(\frac{m_{3/2}}{10 \gev} \right)^{-3}.
\label{lifetime25}}
If
$|U_{\widetilde{\gamma}\nu}|^{2}\sim 10^{-16}-10^{-12}$ in order to reproduce neutrino masses, as we will show below, the gravitino will be very long lived 
as expected (recall that the lifetime
of the Universe is about $10^{17}$ s).

For the gravitino to be a good dark matter candidate we still need to check that it can be present in the right amount to explain the relic density
inferred by WMAP, $\Omega_{DM} h^2 \simeq 0.1$ \cite{wmap}.
% If the R-parity is slightly broken and 
% the Lightest Supersymmetric Particle (LSP) is extremely weakly interacting
% so that the decay rate is sufficiently suppressed 
% then the lifetime of LSP is much longer than the age of our Universe
% and it can be a good candidate of dark matter.
% This is the case for Gravitino dark matter, since the interaction is
% suppressed by the Planck scale.
With the introduction of inflation, the primordial gravitinos are diluted
during the exponential expansion of the Universe. Nevertheless, after inflation,
in the reheating process, the gravitinos are reproduced again from the
relativistic particles in the thermal bath. The yield of gravitinos
from the scatterings is proportional to the reheating temperature, $\treh$, and
estimated to be~\cite{Bolz:2000fu}
\begin{equation}
\Omega_{3/2} h^2 \simeq 0.27\left(\frac{\treh}{10^{10}\gev} \right)
\left(\frac{100\gev}{m_{3/2}} \right)\left(\frac{\mgluino}{1\tev}\right)^2,
\label{oh2buchmuller}
\end{equation}
where $\mgluino$ is the gluino mass. As is well known, adjusting the
reheating temperature one can reproduce the correct relic density for each possible value of the gravitino mass\footnote{ 
Let us recall that there is
a lower limit of 1.2 keV on the mass of (warm) dark matter particles
from Lyman $\alpha$ forest~\cite{Viel:2005qj}.}. 
For example for $m_{3/2}$ of the order of 1--1000 GeV
%, as expected 
%from supergravity scenarios, 
one obtains $\Omega_{3/2} h^2 \simeq 0.1$ for
$T_R\sim 10^8-10^{11}$ GeV, with $\mgluino\sim 1$ TeV.
Even with a high value of $T_R$ there is no gravitino problem, since the
next-to-LSP decays to standard model particles much earlier than BBN epoch
via R-parity breaking interactions.

%Before carrying out the numerical analysis to show that
Let us now show that
$|U_{\widetilde{\gamma}\nu}|^{2}\sim 10^{-16}-10^{-12}$ in the $\mu\nu$SSM.
We can easily make an estimation. For a $2\times2$ matrix,
\begin{equation}
\left(\begin{array}{cc}
a & c\\
c & b\end{array}\right),
\end{equation}
the mixing angle is given by $\tan2\theta=2c/(a-b)$.
% \[
% \tan2\theta=\frac{c}{a-b}.\]
In our case (see Eq.~(\ref{matrizse})) $c\sim g_{1}\nu\sim 10^{-4}$ GeV (represents the mixing of Bino and left handed
neutrino), $a\sim1$ TeV (represents the Bino mass $M_{1}$), and $b=0$.
Thus one obtains $\tan2\theta\sim 10^{-7}$, implying
% \[
% \tan2\theta=\frac{c}{a-b}=\frac{10^{-4}}{10^{3}}\sim10^{-7}.\]
%Since $\theta$ is small 
$\sin\theta\sim\theta\sim10^{-7}$.
This gives $|U_{\widetilde{\gamma}\nu}|^{2}\sim10^{-14}$.
% \[
% |U_{\widetilde{\gamma}\nu}|^{2}\sim10^{-14}.\]
More general, $\theta\sim\frac{g_{1}\nu}{M_{1}}\sim10^{-6}-10^{-8}$, giving rise to
\begin{equation}
%\theta\sim\frac{g_{1}\nu}{M_{1}}\sim10^{-6} \; \text{to}\; 10^{-8} \:\;\;\;\; 
10^{-16} \lesssim |U_{\widetilde{\gamma}\nu}|^{2} \lesssim 10^{-12} \;.
\label{range}
\end{equation}
% We have carried out the numerical analysis of the whole parameter space 
% discussed in Sect. \ref{parameters}, and the results confirm this estimation.
% {\bf Some figures from numerical calculation.}.
% 
% \begin{itemize}
% \item The plot of $\left| U_{\photino \nu}  \right|^2$ with the several parameters.
% \end{itemize}
% 

\begin{figure}[!t]
  \begin{center}
  \begin{tabular}{c c}
   \includegraphics[width=0.6\textwidth]{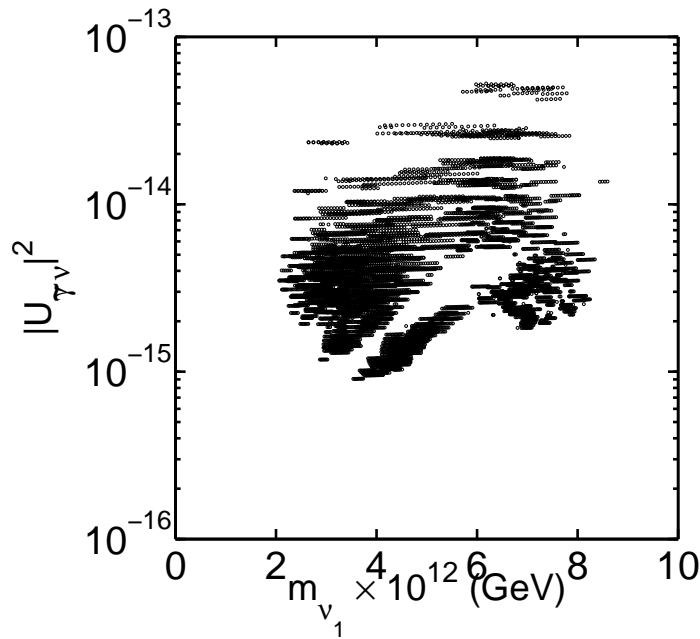} 
&
   \end{tabular}
  \end{center}
 \caption{$|U_{\widetilde{\gamma}\nu}|^{2}$ versus the lightest neutrino mass.} \label{scanplot} 
%For the range $ 0 <  \lambda <  0.4, \; 0 <  \kappa    <  0.4, \;  100 < \nu^c < 3000 \, GeV, \; -3000 \, GeV < M_2 < 0, \;  2 < \tan(\beta) < 40, \; 10^{-7} < \nu_1 < 10^{-5}, \;  10^{-6} < \nu_2 = \nu_3 < 10^{-4}, \;  10^{-7} < Y_{\nu_1} < 10^{-6},\; 10^{-7} < Y_{\nu_2}= Y_{\nu_3} < 10^{-6}. $} \label{scanplot}
\end{figure}

In order to confirm this estimation we have performed a scan 
of the low-energy parameter space of the model discussed in Sect. \ref{parameters}, over the following 
%parameters and 
ranges:
\begin{eqnarray}
0 \leq & \lambda & \leq 0.4, \nonumber \\
0 \leq & \kappa  & \leq 0.4, \nonumber  \\
100 \, \GeV \leq  & \nu^c  & \leq 3 \, \TeV, \nonumber \\
-3 \, \TeV \leq &  M_2  & \leq  0 \, \GeV,  \nonumber \\
2  \leq  & \tan\beta & \leq  40,  \nonumber \\
10^{-7} \,  \GeV  \leq  & \nu_1  & \leq  10^{-5} \,  \GeV,  \nonumber \\
10^{-6} \,  \GeV  \leq  & \nu_2 = \nu_3 & \leq  10^{-4} \, \GeV,  \nonumber \\
10^{-7}  \leq  & Y_{\nu_1}      &  \leq  10^{-6}, \nonumber  \\
10^{-7}  \leq  & Y_{\nu_2}= Y_{\nu_3} & \leq  10^{-6}.
\end{eqnarray}
Concerning the rest of the soft parameters, we will take for simplicity in the computation
$m_{\tilde Q,\tilde u^c,\tilde d^c,\tilde e^c}=1$ TeV,
$A_{u,d,e}=1$ TeV, $A_{\lambda}=-A_{\nu}=-2A_{\kappa}=1$ TeV, and for the other gaugino masses we will use the
GUT relations. 
% $M_1 = \frac{\alpha_1^2}{\alpha_2^2} M_2$,
% $M_3 =  \frac{\alpha_3^2}{\alpha_2^2} M_2$,
% implying $M_1\approx 0.5 M_2$, $M_3\approx 2.7 M_2$.
Although this is not a full exploration of the parameter space, which 
is beyond the scope of this work, it gives a fair estimation of the 
representative values for $|U_{\widetilde{\gamma}\nu}|^2$. 
The results are shown in Fig. \ref{scanplot}.
The black points there correspond to regions of the parameter space where
the current data on neutrino masses and mixing angles are reproduced
(where we are using the allowed 3$\sigma$ ranges discussed in \cite{neuranges}).
In addition, these regions avoid false minima and tachyons, as well as fulfil the 
Landau pole constraint, following the lines discussed in \cite{MuNuSSM2,neutrinos}.
Typically, the mass of the lightest neutralino is above 20 GeV, and since the gravitino mass in this model is constrained to be below that value, as we will see in the next section, the gravitino can be used as the LSP.

In principle, we could conclude that the range 
$10^{-15} \lesssim |U_{\widetilde{\gamma}\nu}|^{2} \lesssim 5 \times 10^{-14}$ 
is specially favoured. 
Nevertheless, despite that we see only a few solutions for 
$|U_{\widetilde{\gamma}\nu}|^{2}<10^{-15}$, looking at Eq. (\ref{umix})
we could infer that values close to zero would be achievable 
through a cancellation of the Bino and Wino contribution. 
Therefore we consider that a good estimation for the lower bound  
of $|U_{\widetilde{\gamma}\nu}|^2$ without much fine tunning is $10^{-16}$.
On the other hand, one could get values 
$|U_{\widetilde{\gamma}\nu}|^{2} > 10^{-13}$ in the regime of degenerated neutrinos, where larger values of the lightest neutrino mass than the ones shown 
in Fig. \ref{scanplot} are required (i.e. allowing an exploration for 
larger values of $Y_{\nu_i}$ and the VEVs $\nu_i$). 
In this sense, one could use the (conservative) range written in Eq. (\ref{range}).  
This is what we will do in the next section.
%, although for discussing specific examples we will use 
%values for $|U_{\widetilde{\gamma}\nu}|^{2}$ within the range
%$10^{-15} \lesssim |U_{\widetilde{\gamma}\nu}|^{2} \lesssim 5 \times 10^{-14}$, %specially favoured by our scan.

%In the following, and since 
% From Fig. (\ref{scanplot}), we could conclude that the range between 
% $10^{-15} \lesssim |U_{\widetilde{\gamma}\nu}|^{2} \lesssim 3 \times 10^{-14}$ 
% is specially favoured and using the arguments given above and being 
% conservative, the representative values of 
% $ |U_{\widetilde{\gamma}\nu}|^{2}$ for this model are within the range:
% \begin{equation}
%\theta\sim\frac{g_{1}\nu}{M_{1}}\sim10^{-6} \; \text{to}\; 10^{-8} \:\;\;\;\; 
% 10^{-16} \lesssim |U_{\widetilde{\gamma}\nu}|^{2} \lesssim 10^{-12} \;,
% \end{equation}
%but since 
% the range $10^{-15} \lesssim |U_{\widetilde{\gamma}\nu}|^{2} \lesssim 3 \times 10^{-14}$ is specially favoured by our scan, we will use 
% values for $|U_{\widetilde{\gamma}\nu}|^{2}$ within this range. 
%in the next section.
% A full exploration of the parameter space is out of the scope of this work.  The estimation above presented  is fundamented in the scan showed in fig. \ref{umix}, that is not complete but enough for our propose, and several consideration above described that include the cases that could scape such incomplete scan.

\section{Gamma rays from gravitino decay}

%\subsection{Gamma ray from the decay of Dark Matter}

Since in R-parity breaking models the gravitino decays producing a monochromatic photon with an energy $m_{3/2}/2$, one can try to extract constraints on the parameter space
from gamma-ray observations~\cite{Takayama:2000uz,Overduin:2004sz}.
Actually, model independent constraints on late dark matter decays
using the gamma rays were studied in~\cite{Yuksel:2007dr} (see also also~\cite{PalomaresRuiz:2007ry} for the case of neutrino production).
{\bf \Large}
\begin{figure}[!t]
  \begin{center}
  \begin{tabular}{c c}
   \includegraphics[width=0.5\textwidth]{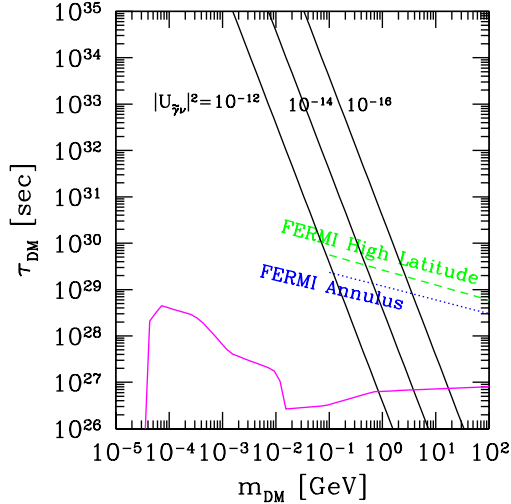} 
&
   \end{tabular}
  \end{center}
 \caption{Constraints on lifetime versus mass for
a decaying dark matter particle. 
The region below the magenta solid line is excluded by gamma-ray observations \cite{Yuksel:2007dr}. 
The region below the green dashed (blue dotted) line will be checked by FERMI.
Black solid lines correspond to the predictions of the $\mu\nu$SSM for
several representatives values of 
$|U_{\widetilde{\gamma}\nu}|^{2}=10^{-16}-10^{-12}$.} \label{ExclusionPlot}
\end{figure}
There, the decaying dark matter was constrained using the gamma-ray line emission limits from the galactic center region obtained with the SPI spectrometer 
on INTEGRAL satellite, and the isotropic diffuse photon background as determined from SPI, COMPTEL and EGRET data.
These constraints are shown with a magenta line in Fig.~\ref{ExclusionPlot}, where
a conservative non-singular profile at the galactic center is used.
%and an anisotropoic sharp line from the halo of our galaxy~\cite{Asaka:1998ju}

On the other hand, the FERMI satellite~\cite{Fermi0} launched in June 2008 is able to 
measure gamma rays with energies between 0.1 and 300 GeV.
We also show in Fig.~\ref{ExclusionPlot} 
the detectability of FERMI 
in the 'annulus' and 'high latitude' regions
following the work in~\cite{bertone}.
Below the lines, FERMI will be able to detect the signal from decaying dark matter.
Obviously, no signal means that the region would be excluded and FERMI would have
been used to constrain the decay of dark matter~\cite{bertone}.

% Now Fermi-LAT showed that there is no access around $10 \gev$ which was 
% claimed by EGRET data and very close to the conventional model.
% This put the stronger bound on the additional Gamma ray source than 
% using EGRET data as background. We will examine the implication of
% $\mu\nu$SSM in gamma ray observation.
% If the Fermi-LAT does not see any differences then the observation can be used 
% to constrain the decay of dark matter~\cite{bertone}.

Finally, we show in the figure with black solid lines the values of the parameters predicted by the $\mu\nu$SSM using Eq.~(\ref{lifetime25}), for
several representative values of 
$|U_{\widetilde{\gamma}\nu}|^{2}$.
%=10^{-16}-10^{-12}$.
We can see that values of the gravitino mass larger than 20 GeV are disfavored in this model by the isotropic diffuse photon background observations 
(magenta line).
In addition, FERMI will be able to check important regions of the parameter 
space with gravitino mass between $0.1 - 20$ GeV and $|U_{\widetilde{\gamma}\nu}|^{2}=10^{-16}-10^{-12}$ (those below the green line).

Let us now discuss in more detail what kind of signal is expected to be observed by FERMI
if the gravitino lifetime and mass in the $\mu\nu$SSM (black solid lines) correspond to a point below the
green line 
%({\bf OR OF THE LINE CORRESPONDING TO MID LATITUDE IN CASE WE DRAW IT})
in Fig.~\ref{ExclusionPlot}

% We consider the line spectrum of photons when gravitino decay to photon 
% and neutrino by two-body decays.
% If the gravitino is sufficiently heavy then it can decay into $W$ and $Z$
% bosons. However considering the predicted mixing between photino and neutrino
% $\left| U_{\photino \nu}  \right|$ in the $\mu\nu$SSM, 
% the Gravitino heavier than around $20\gev$
% is already ruled out by the EGRET data~\cite{Yuksel:2007dr}.
% Since the EGRET observation of gamma ray is much higher than the 
% background, when we use the new data from Fermi-LAT which is almost close 
% to the conventional background the constraint on decaying dark matter is
% much stronger.

As it is well known, there are two sources for a diffuse background from dark matter decay.
One is the cosmological diffuse gamma ray coming from extragalactic regions, and the other is the one coming from the halo of our galaxy.

The photons from cosmological distances are red-shifted during their journey to 
the observer and the isotropic extragalactic flux turns out to be~\cite{Takayama:2000uz,Overduin:2004sz,bertone} 
\dis{
\frac{d J_{eg}}{dE}= A_{eg}\frac{2}{\mdm}\left(1+\kappa\bfrac{2E}{\mdm}^3  \right)^{-1/2} \bfrac{2E}{\mdm}^{1/2} \Theta\left(1-\frac{2E}{\mdm}  \right),
}
with 
\dis{
A_{eg}=\frac{\Omega_{DM}\rho_c}{4\pi\taudm \mdm H_0\Omega_M^{1/2} }=2.11\times 
10^{-7} (\cm^2\ \text{s}\ \str)^{-1} \bfrac{\taudm}{10^{27}\text{s}}^{-1}\bfrac{\mdm}{10\gev}^{-1}.
}
Here $\kappa=\Omega_\Lambda/\Omega_M\simeq 3$
with $\Omega_\Lambda+\Omega_M=1$, $\rho_c=1.05\ {h^2}\times 10^{-5} \gev\cm^{-3}$, $H_0=h\ 100 \km\ s^{-1}\ \mpc^{-1}$ with $h=0.73$~\cite{Yao:2006px}, and
$\taudm$ and $\mdm$ are the lifetime and mass of the dark matter particle, respectively. We take the dark matter density as $\Omega_{DM} h^2=0.1$.

On the other hand, 
the photon flux from the galactic halo shows an anisotropic sharp line.
For decaying dark matter this is given by
\dis{
\frac{d J_{halo}}{dE}= A_{halo}\frac{2}{\mdm}\delta\left(1-\frac{2E}{m_{DM}} \right),
}
with 
\dis{
A_{halo}= \frac{1}{4\pi\taudm\mdm}\int_{\textrm{los}}\rho_{halo}(\vec{l})d\vec{l}\ ,
}
where the halo dark matter density is integrated along the line of sight, and
we will use a NFW profile \cite{navarro}
\dis{
\rho_{NFW}(r)=\frac{\rho_h}{r/r_c(1+r/r_c)^2}\ ,
}
where we take $\rho_h=0.33 \gev/\cm^3=0.6\times10^5 \rho_c$, $r_c=20$ kpc, and
$r$ is the distance from the center of the galaxy.
% with $\rho_h=0.33 \gev/\cm^3=0.6\times10^5 \rho_c$ and $r_c=20$ kpc.
%However our results will not depend much on this halo profile,
%since we will use the region of the sky away from the Galactic center and 
%Galactic plane.
The latter 
can be re-expressed using the distance from the Sun, $s$, in
units of $\Rsun=8.5$ kpc (the distance between the Sun and the galactic center) and
the galactic coordinates, the longitude, $l$, and the latitude, $b$, as
\dis{
r^2(s,b,l)=\Rsun^2[(s-\cos b\cos l)^2+(1-\cos^2b \cos^2l)]\ .
}

\begin{figure}[!t]
%  \begin{center}
  \begin{tabular}{c c}
   \includegraphics[width=0.5\textwidth]{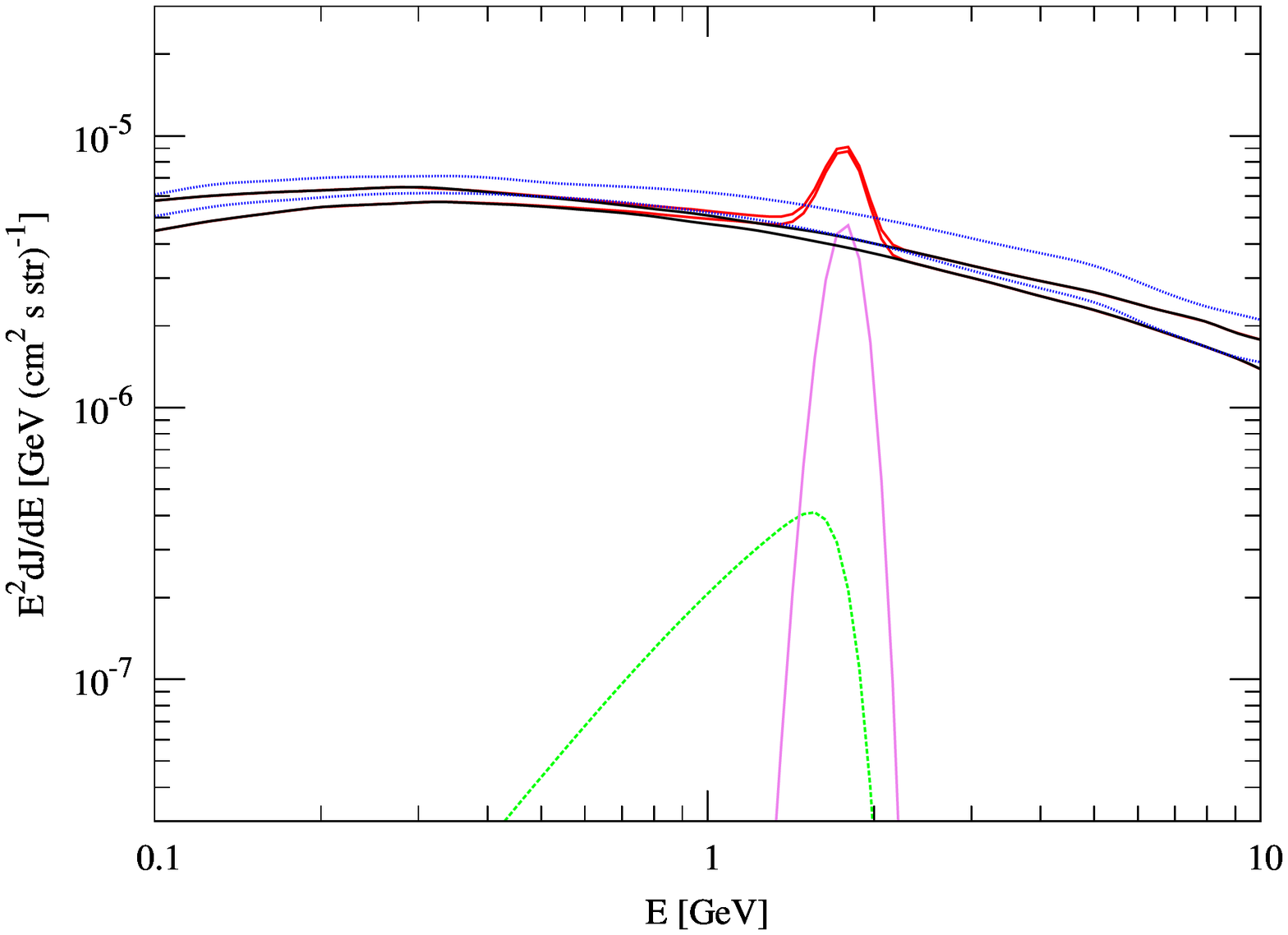}
&
   \includegraphics[width=0.5\textwidth]{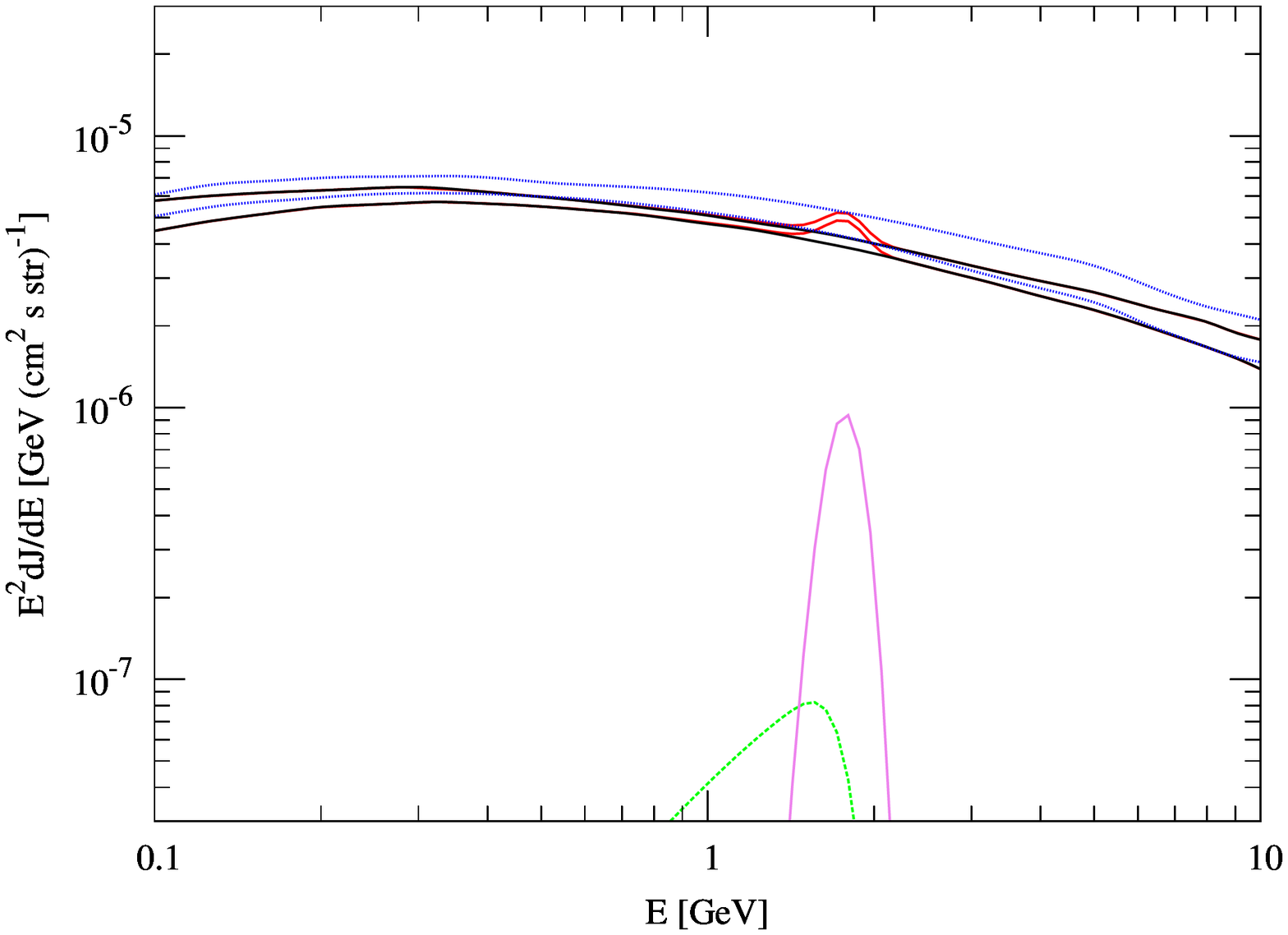}
      \\ (a)& (b)\\
   \end{tabular}
%  \end{center}
 \caption{
Expected gamma-ray spectrum 
for an example of gravitino dark matter decay 
in the mid-latitude range ($10^{\, \circ}\le |b|\le 20^{\, \circ}$) in the $\mu\nu$SSM
with 
$m_{3/2}=3.5\gev$ and (a)
$\left| U_{\photino \nu}  \right|^2= 8.8\times10^{-15}$ corresponding to  
$\tau_{3/2}=10^{27}$ s, (b) 
$\left| U_{\photino \nu}  \right|^2= 1.7\times10^{-15}$ corresponding to  
$\tau_{3/2}=5\times10^{27}$ s.
The 
green dashed, magenta solid, and black solid lines correspond to the diffuse
extragalactic gamma ray flux, the gamma-ray flux from the halo, and to the
conventional background, respectively.
The total gamma-ray flux is shown with red solid lines.
The blue solid lines are explained in the note added in Sect. 6.
} \label{Gravitinodecay}
\end{figure}

\begin{figure}[!t]
  \begin{center}
  \begin{tabular}{c c}
   \includegraphics[width=0.5\textwidth]{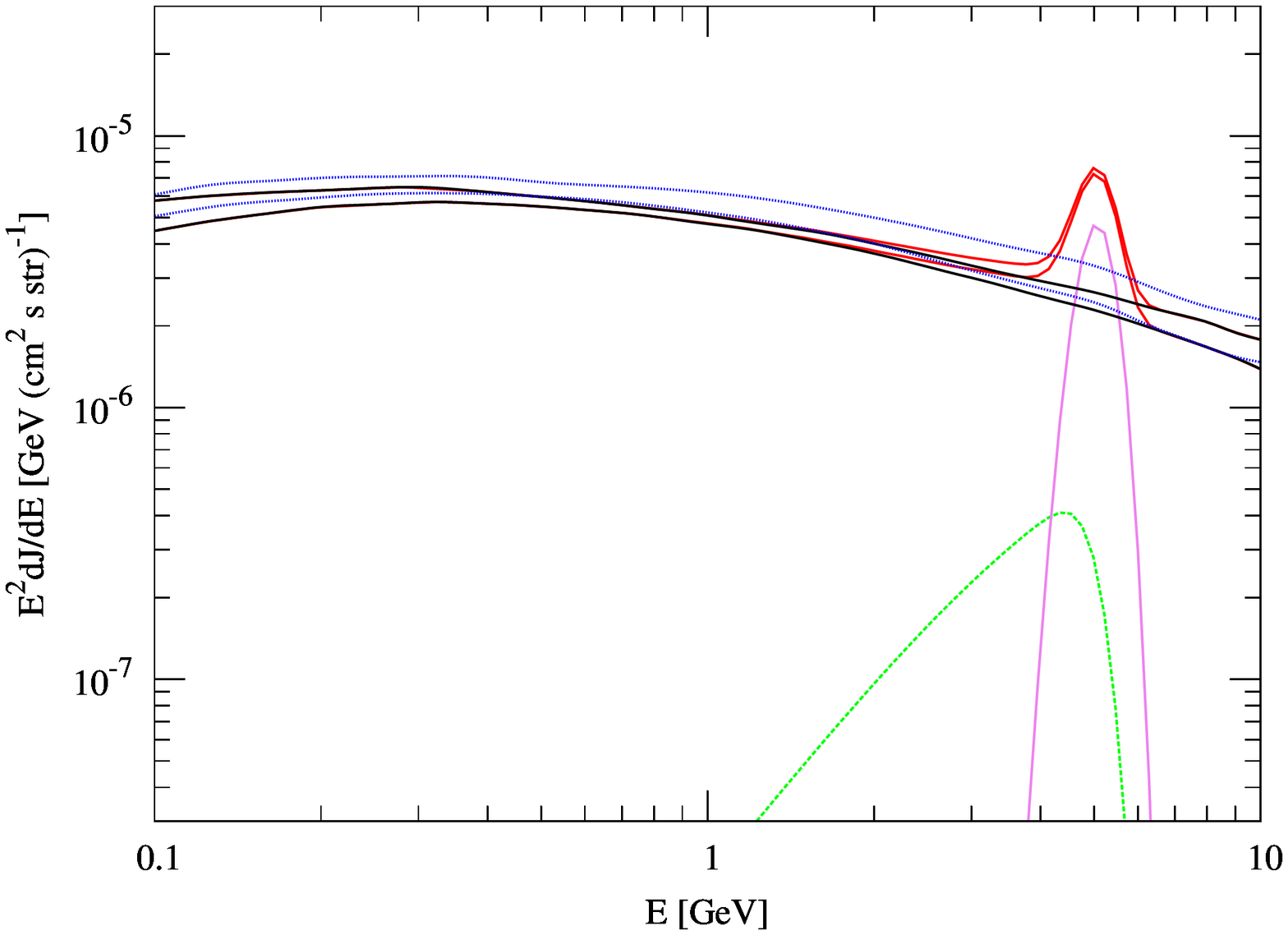}
&
  \includegraphics[width=0.5\textwidth]{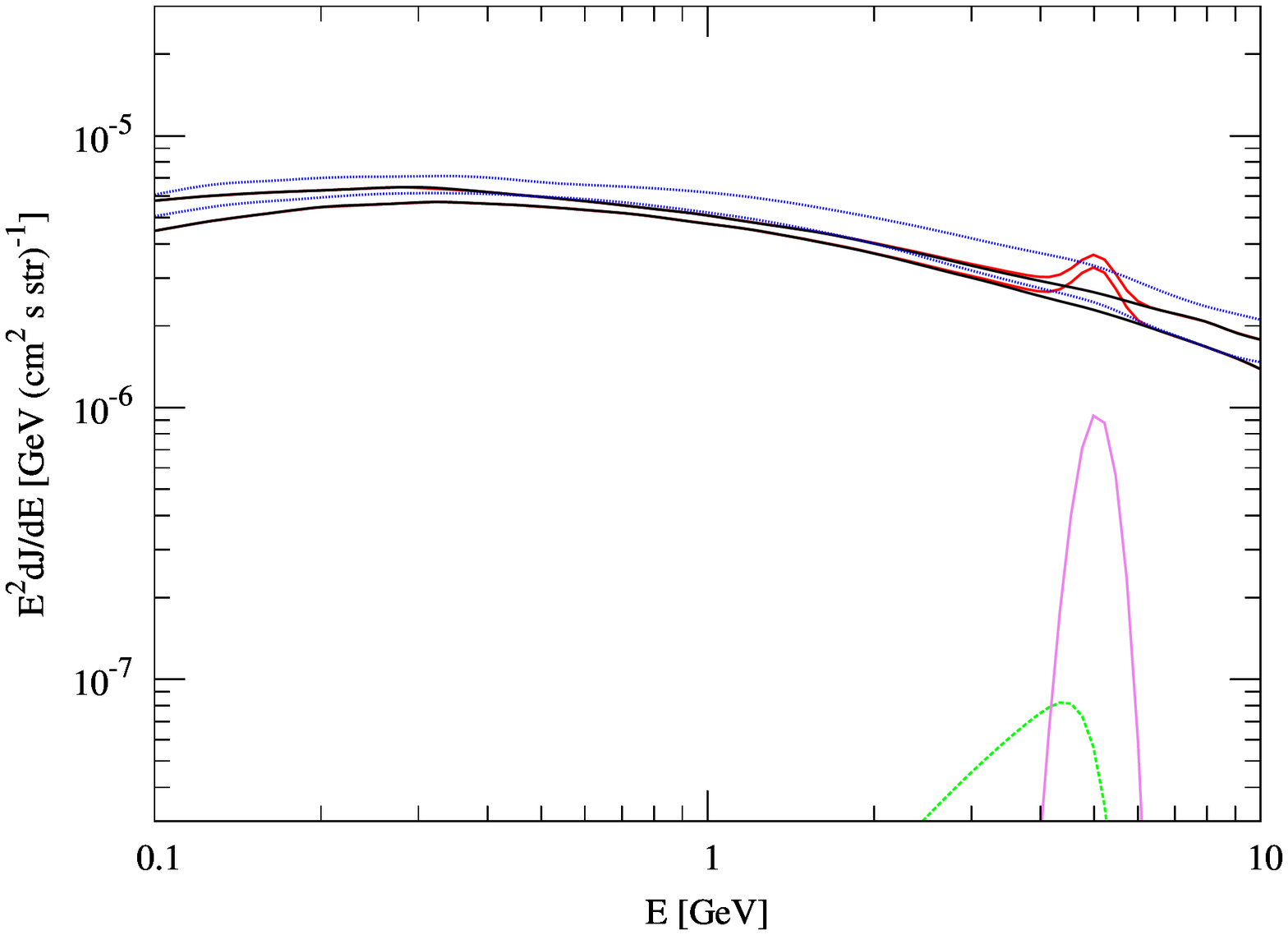}\\
 (a)& (b)\\
   \end{tabular}
  \end{center}
 \caption{
The same as in Fig. \ref{Gravitinodecay} but for 
$m_{3/2}=10$ GeV and (a)
$\left| U_{\photino \nu}  \right|^2= 3.8\times 10^{-16}$
corresponding to  
$\tau_{3/2}=10^{27}$ s, (b) 
$\left| U_{\photino \nu}  \right|^2= 7.6\times 10^{-17} $
corresponding to  
$\tau_{3/2}=5\times10^{27}$ s.
} \label{Gravitinodecay2}
\end{figure}

It is worth noticing here that when computing above the gamma-ray fluxes, the effects
of attenuation of the flux in the interstellar or the intergalactic medium have been 
neglected. In our mass range of the gravitino, the flux from the decay of the gravitino dark matter is made of photons and neutrinos, thus we might
expect the attenuation of the gamma-ray flux by pair production. Nevertheless, for our case with less than 10 GeV gamma-ray flux the attenuation is suppressed both in the
galactic and extragalactic medium \cite{moska}, and can therefore 
be safely neglected.

Let us now compute with the above formula, as an example, the expected diffuse gamma-ray
emission in the mid-latitude range ($10^{\, \circ}\le |b|\le 20^{\,\circ}$),
which is being analized by FERMI \cite{morselli},
for the case of gravitino dark matter.
Let us assume for instance a value of $m_{3/2}=3.5\gev$ and
$\left| U_{\photino \nu}  \right|^2= 8.8\times10^{-15}\, (1.7\times10^{-15})$ in the $\mu\nu$SSM, corresponding to
$\tau_{3/2}=10^{27}$ ($5\times10^{27}$) s, using Fig.~\ref{ExclusionPlot}.
We convolve the signal with a Gaussian distribution with the energy resolution  $\Delta E/E=0.08$, between $E=1-10\gev$, following~\cite{Fermi0}, and  
then we average the halo signal over the region for the mid-latitude range mentioned above.

The results for the two examples are shown in Fig.~\ref{Gravitinodecay}. There, the green dashed line corresponds to the diffuse extragalactic gamma ray flux,
and the magenta solid line corresponds to the gamma-ray flux from the halo.
The black solid lines represent the background including the diffuse galactic emission model from GALPROP \cite{galprop}, and point source and isotropic 
contributions \cite{morselli}. The systematic uncertainties for the latter
generate the band shown within the two black lines.

The total gamma-ray flux,
including background, extragalactic, and line signal, is shown with red solid lines.
%Notice that in this example we have choosen the parameters of the $\mu\nu$SSM
%in such a way that 
We can see that
% The observation almost agree with the conventional background model,
% except slight excess around $3\gev$.
% As an example of the flux from Gravitino decay we show figure~\ref{Gravitinodecay} with
% $\mgravitino=3.5\gev$ and $\tau_\gravitino=8\times10^{26}\sec$.
% We convolved the signal with a Gaussian distribution with the energy resolution  %$\Delta E/E=0.09$  between $E=1-10\gev$ following Fermi-LAT~\cite{Fermi0}. 
% Then we averaged the halo signal over a region for mid-latitude range ($0^o\le l \le 360^o, 10^o\le |b|\le 20^o$)~\cite{slide}. 
% Here the background (black lines) and Fermi observation (magenta points) are 
% also taken from that slide.
% The magenta sharp line is the gamma ray flux from the halo of our galaxy from Gravitino decay. The green dashed line is the diffuse extragalactic gamma ray from Gravitino decay. The red lines are the total flux of gamma ray including the background, extragalactic gamma ray and line signal from Gravitino decay.
the sharp line signal associated to an energy half of the gravitino mass, dominates the extragalactic signal
and can be a direct measurement (or exclusion) in the FERMI gamma ray observation.
We could also use the EGRET data \cite{egret} to constrain the parameter
space of the model. 
Although the data beyond 1 GeV are controversial (even instrumental effects 
might be a possible explanation for the observed excess \cite{instrumental}),
the line obtained in the example of Fig.~\ref{Gravitinodecay}a 
is very sharp and could be discarded when compared with the spectrum
of EGRET.

As another example, we show in Fig.~\ref{Gravitinodecay2}
the case of
$m_{3/2}=10$ GeV for the same values of lifetimes as in Fig.~\ref{Gravitinodecay}.

\section{Conclusions}

We have discussed the possibility of gravitino dark matter in the 
$\mu\nu$SSM, where R-parity is broken and therefore the LSP is unstable.
For the gravitino mass of the order of $\gev$, the lifetime of the gravitino is much longer than the age of the Universe
and the observed relic density can be explained well by the thermal production of gravitinos after inflation.
%and the reheating temperature can be adjusted to reproduce the observed relic density.
As a consequence, the gravitino can be a good candidate for dark matter.

We have also studied the prospects for detecting gamma rays from decaying 
gravitinos.
If the gravitino explains the whole dark matter component,
the gravitino mass larger than $20\gev$ is disfavored by 
%in the  $\mu\nu$SSM using 
the isotropic diffuse photon background measurements. 
Nevertheless, a gravitino with a mass range between $0.1 - 20$ GeV gives rise to 
a signal that might be observed by the FERMI satellite. 
In this way important regions of the parameter space of the $\mu\nu$SSM can be
checked.
%important regions of the parameter space of the $\mu\nu$SSM can be checked in this way.

{\bf \Large}
\begin{figure}[!t]
  \begin{center}
  \begin{tabular}{c c}
   \includegraphics[width=0.5\textwidth]{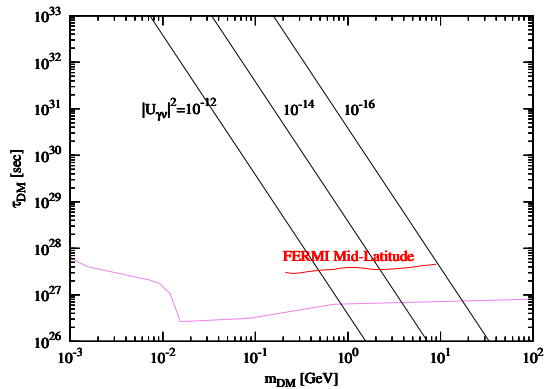} 
&
   \end{tabular}
  \end{center}
 \caption{Constraints on lifetime versus mass for gravitino dark matter 
in the $\mu\nu$SSM. 
The region below the magenta solid line is excluded by several gamma-ray observations \cite{Yuksel:2007dr}. 
The region below the red solid line is disfavoured by FERMI.
Black solid lines correspond to the predictions of the $\mu\nu$SSM for
several representatives values of 
$|U_{\widetilde{\gamma}\nu}|^{2}=10^{-16}-10^{-12}$.} \label{ExclusionPlot333}
\end{figure}

\section{Note added}

After completion of the current work, 
the FERMI experiment reported 5-month measurements of the diffuse gamma-ray emission
in the mid-latitude range \cite{ferminew}.
We have added in Figs. 3 and 4 blue solid lines corresponding to these Fermi LAT data with systematic
uncertainties.
These turn out to be consistent with the background model, implying that the sharp lines obtained in the examples of Figs. 3a and 4a have not been observed.
Taking these results into account,
we have summarized in Fig. 5 the constraints on lifetime versus mass for the 
$\mu\nu$SSM.
Values of the gravitino mass larger than 10 GeV are now disfavored, as well as
lifetimes smaller than about 3 to 5$\times 10^{27}$ s.

\vspace{0.3cm}

\noindent
{\bf Acknowledgments}

We gratefully acknowledge G. Bertone for his collaboration during the early
stages of this work.
We thank A. Morselli for very helpful information concerning
FERMI. We also thank G.A. G\'omez for his valuable comments.
The work of the authors was supported by the spanish MICINN's Consolider-Ingenio 2010 Programme under grant MULTIDARK CSD2009-00064.
The work of K.Y. Choi and C. Mu\~noz was supported 
in part by MICINN under grants FPA2009-08958 and FPA2009-09017,
by the Comunidad de Madrid under grant HEPHACOS S2009/ESP-1473,
and by the European Union under the Marie Curie-ITN program PITN-GA-2009-237920.
D.E. L\'opez-Fogliani thanks STFC for financial support. 
The work of R. Ruiz de Austri was supported in part by MICINN under grant
FPA2007-60323, by the Generalitat Valenciana under grant PROMETEO/2008/069,
and by the Spanish Consolider-Ingenio 2010 Programme CPAN (CSD2007-00042).
We also
thank the ENTApP Network of the ILIAS project RII3-CT-2004-506222 
and the UniverseNet Network MRTN-CT-2006-035863.

\end{document}